\documentstyle[aps,prb,twocolumn]{revtex}

\begin{document}
\title{Fractional exclusion statistics and shot noise in ballistic conductors}
\author{G. Gomila and L. Reggiani}
\address{Dipartimento di Ingegneria dell' Innovazione and \\
Istituto Nazionale di Fisica della Materia \\
Universit\'a di Lecce,\\
Via Arnesano s/n, 73100-Lecce, Italy}
\maketitle

\begin{abstract}
We study the noise properties of ballistic conductors with carriers
satisfying fractional exclusion statistics. To test directly the nature of
exclusion statistics we found that systems under weakly degenerate
conditions should be considered. Typical of these systems is that the
chemical potential, $\mu $ is in the thermal range $|\mu |<3k_{B}T$. In
these conditions the noise properties under current saturation are found to
depend upon the statistical parameter $g$, displaying suppressed shot noise
for $1/2\leq g\leq 1$, and enhanced shot noise for $0<g<1/2$, according to
the attractive or repulsive nature of the carrier exclusion statistics.
\end{abstract}



\section{Introduction}

The concept of fractional exclusion statistics (FES) was introduced by
Haldane\cite{Haldane91} as a phenomenological description of excitations
with mixed statistical properties (intermediate between fermions and bosons)
in strongly correlated many-body systems. FES is based on the assumption
that the change of the number of available one-particle states in a system,
for a given volume and with fixed boundary conditions, $\Delta D$, depends
linearly on the change of the number of quasiparticles $\Delta N$, i.e. $%
\Delta D=-g\Delta N$. Here, $g$ is the statistical parameter with the
properties of: (i) being independent of the number of quasiparticles and,
(ii) being determined solely by the interaction strength between particles.
In spite of its phenomenological ground, FES has been shown to be realized
by, or directly related to, several physical models, as for instance: (i)
models for strongly correlated particles in one dimension\cite
{Isakov94a,Ha94,Murthy94,Wu95,Hatsugai96,Pepin99} and two dimensions,\cite
{Bhaduri96,Srivastava99} (ii) anyons \cite{Dasnieres94,Chen95} and, (iii)
quasi-particle excitations in the fractional quantum Hall effect (FQHE)
systems.\cite{Su96,Yu97,Wu98,Elburg98}

The quantum statistical mechanics of a generalized gas of particles obeying
FES was pioneered in Refs.[\onlinecite{Dasnieres94,Isakov94b,Wu94}]. Since
then, a series of works were devoted to the thermodynamic properties of this
generalized system.\cite{Rajagopal95,Isakov96,Isakov97,Iguchi97,Su98} In
addition, transport properties have been widely investigated. The linear
response of one dimensional systems obeying FES has been addressed by means
of the Landauer approach\cite{Rego99,Krive99,Isakov99} and the correlation
function method.\cite{Greiner99} The non linear response for edge
excitations in the FQHE in terms of FES has been also analyzed.\cite
{Elburg98}

By contrast, the fluctuation properties of systems satisfying FES have
received only a minor consideration. A significative attempt to this subject
was presented in Ref.[\onlinecite{Isakov99}] where a generalization of the
first quantized Landauer approach to conductance and noise was proposed for
particles obeying FES. The application of the formalism to the case of
strongly degenerate systems revealed that the effects of carrier statistics
on the noise properties is only appreciated in the transition region between
thermal and shot noise. Surprisingly enough, the shot noise suppression
factor was found to be independent of the particle statistics. Thus, no
evidence of the ''attractive'' or ''repulsive'' nature of the different
statistics was expected.

The aim of the present paper is, precisely, to analyze under which
conditions the noise properties of systems obeying FES are expected to
depend on the statistics of the carriers, thus shedding new light on the
attractive or repulsive nature of FES. To this purpose, we investigate the
noise properties of a one dimensional ballistic system obeying FES with
current voltage ($I-V$) characteristics ranging from linear to current
saturation regimes. As prototype of a one dimensional ballistic system, we
consider a perfectly transmitting channel where the random injection of
carriers from the contacts is the only source of noise. Being related to the
occupation probabilities, this noise source is only present at temperatures
different from zero and it depends on the carrier statistics. Therefore, the
above prototype represents the simplest system where to study the
inter-relations between FES and noise.

The content of the paper is organized as follows. In Sec. \ref{System} we
detail the system under study. In Sec. \ref{Degenerate} we analyze the
transport and noise properties for the case of strongly degenerate
conditions. In Sec. \ref{Weakly} the case of weakly degenerate conditions
are considered. Finally, in Sec. \ref{Conclusions} we sum up the main
conclusions of the paper.

\section{System under study}

\label{System} We consider a standard two terminal experiment where two
reservoirs are adiabatically inter-connected by a one-dimensional channel
with the following assumptions. (i) The channel is ballistic and perfectly
coupled with the reservoirs acting as ideal contacts so that no reflections
take place at the interfaces. (ii) Each reservoir is in quasi-equilibrium at
a temperature $T$, and electrochemical potential $\phi $. (iii) All the band
bending occurs in the channel and the relative position of the conduction
band and the electrochemical potential does not change in the contacts, i.e.
the chemical potential $\mu =\phi -E_{C}$, with $E_{C}$ the bottom of the
conduction energy band, is independent of the applied bias (for simplicity
we assume the same value for the two contacts). Therefore, when the bias $%
qV=\phi _{2}-\phi _{1}$, with $-q$ the carriers charge (taken as negative),
is changed, the potential can vary exclusively inside the ballistic channel,
and the contacts can be excluded from consideration.

The carriers in the system are assumed to satisfy FES and are injected from
the reservoirs into the channel in accordance with the corresponding
equilibrium distribution function. For a generalized gas satisfying FES,
standard quantum statistics shows that its equilibrium distribution
function, $f_{g}(\varepsilon -\mu )$, satisfies the implicit equation\cite
{Dasnieres94,Isakov94b,Wu94} 
\begin{equation}
\left( 1-gf_{g}(z)\right) ^{g}\left( 1+(1-g)f_{g}(z)\right)
^{1-g}=f_{g}(z)e^{\frac{z}{k_{B}T}}\text{,}  \label{f_alfa}
\end{equation}
where $\varepsilon $ is the carrier energy, $k_{B}$ the Boltzmann constant
and $T$ the temperature. The limiting values $g=1$ and $g=0$ correspond to
the Fermi-Dirac and Bose-Einstein statistics, respectively. Valuable
approximations of the above distributions are: (i) under strongly degenerate
conditions $f_{g}(\varepsilon -\mu )=(1/g)\theta (\mu -\varepsilon )$ with $%
\theta $ the Heaviside function, (ii) under non-degenerate conditions, $%
f_{g}(\varepsilon -\mu )=\exp (-(\varepsilon -\mu )/k_{B}T)$, which being
the Maxwell-Boltzman distribution is independent of $g$ . Here, for the sake
of convenience we will take $\mu >3k_{B}T$ and $\mu <-3k_{B}T$ as synonymous
of strongly degenerate and non-degenerate conditions, respectively. These
conditions are well verified for all FES. Finally, we note that the charge
of carriers, $q$, is here taken to be independent of carrier statistics to
allow us focusing only on purely statistical effects. To include the
combined effect of fractional charge and fractional statistics one only
needs to substitute $q$ by $ge$ in the final results.

\section{Strongly degenerate systems}

\label{Degenerate} For the sake of simplicity we first consider the case of
strongly degenerate conditions (i.e., $\mu >3k_{B}T$). Under these
conditions, one can neglect long range Coulomb interaction effects on the
fluctuations,\cite{Gomila00} thus simplifying considerably the calculations.

\subsection{Transport properties}

Under the conditions that long range coulomb interaction can be neglected,
the average current flowing through a ballistic one dimensional channel can
be written as:\cite{Rego99,Isakov99} 
\begin{equation}
I=(2s+1)\frac{q}{h}\int_{0}^{+\infty }\left[ f_{g}(\varepsilon -\mu
)-f_{g}(\varepsilon +qV-\mu )\right] d\varepsilon \text{,}  \label{I}
\end{equation}
where $s$ is the spin factor, $h$ is the Planck constant and use is made
that $\phi _{2}-\phi _{1}=qV$. For FES, with the help of the relation 
\begin{equation}
f_{g}^{\prime }(z)=\frac{-1}{k_{B}T}f_{g}(z)\left( 1-gf_{g}(z)\right) \left(
1+(1-g)f_{g}(z)\right) \text{,}  \label{df_alfa}
\end{equation}
which follows from Eq. (\ref{f_alfa}), one can integrate the current
equation obtaining the $I-V$ characteristics in explicit form 
\begin{eqnarray}
I &=&I_{0}\left[ \ln \left( \frac{1+(1-g)f_{g}(-\mu )}{1-gf_{g}(-\mu )}%
\right) \right.  \nonumber \\
&&\left. -\ln \left( \frac{1+(1-g)f_{g}(qV-\mu )}{1-gf_{g}(qV-\mu )}\right)
\right] \text{.}  \label{IV}
\end{eqnarray}
Here, $I_{0}=(2s+1)\frac{q}{h}k_{B}T=G_{0}\frac{k_{B}T}{q}$, with $%
G_{0}=(2s+1)\frac{q^{2}}{h}$.

The $I-V$ characteristics given by Eq. (\ref{IV}) is plotted in Fig. \ref
{I-V} for different statistics and strongly degenerate conditions given by $%
\mu /k_{B}T=8$. In all the cases we found, that the $I-V$ characteristics
are linear for applied potentials lower than the chemical potential, while
they tend to saturate for applied potentials higher than the chemical
potential. The occurrence of current saturation is due to the fact that: (i)
all carriers injected from the contact at higher voltage are collected by
the opposite contact and, (ii) all carriers injected from the contact at
lower voltage return back to the same contact. We anticipate, that the
saturation condition plays a determinant role in evidencing the effects of
the statistics on the noise properties of the system. As can be seen in Fig. 
\ref{I-V}, the main effect of the carrier statistics is to determine the
slope of the linear regime (i.e. the conductance) and the value of the
saturation current. The dependence of these quantities on the statistical
parameter can be obtained straightforwardly. Indeed, from Eq.(\ref{I}) the
(differential) conductance is given by: \cite{Rego99,Krive99,Isakov99} 
\begin{eqnarray}
G &=&\frac{dI}{dV}=G_{0}\int_{0}^{+\infty }\left[ -f_{g}^{\prime
}(\varepsilon +qV-\mu )\right] d\varepsilon  \nonumber \\
&=&G_{0}f_{g}(qV-\mu )\text{,}  \label{G}
\end{eqnarray}
thus yielding for the (linear) conductance (when $qV\ll \mu $) 
\begin{equation}
G^{eq}=G_{0}f_{g}(-\mu )\text{.}  \label{Geq}
\end{equation}
Under strongly degenerate conditions the conductance thus becomes 
\begin{equation}
G^{eq,\deg }=\frac{G_{0}}{g}=(2s+1)\frac{q^{2}}{gh}\text{.}  \label{Geq_deg}
\end{equation}
Therefore, $G^{eq,\deg }$ is found to depend inversely upon the statistical
parameter, in agreement with existing results.\cite
{Rego99,Krive99,Isakov99,Greiner99}

For the saturation current we obtain the expression: 
\begin{eqnarray}
I_{s} &=&(2s+1)\frac{q}{h}\int_{0}^{+\infty }f(\varepsilon -\mu )d\varepsilon
\nonumber \\
&=&I_{0}\ln \left( \frac{1+(1-g)f_{g}(-\mu )}{1-gf_{g}(-\mu )}\right) \text{,%
}  \label{Is}
\end{eqnarray}
which, under strongly degenerate conditions, leads to: 
\begin{equation}
\text{ }I_{s}^{\deg }=G^{eq,\deg }\frac{\mu }{q}=(2s+1)\frac{q\mu }{gh}\text{%
.}  \label{Is_deg}
\end{equation}
The saturation current is a function of the statistical parameter only
through the value of $G^{eq,\deg }$, and thus it also depends inversely upon
the statistical parameter. This dependence is due to the fact that for a
given value of $g$ the occupation number goes as $1/g$ while the velocity of
the injected carries remains the same for all the statistics, once a value
of the chemical potential is given. We remark, that if the unit charge of
carriers depends on the statistical parameter as $q=ge$, then the value of
the saturation current is independent of carrier statistics.

It is worth noting that, a part from the values of the linear conductance
and the saturation current, the effects of the carrier statistics is also
manifested in the shape of the region of transition between the linear and
the saturation regimes. This effect is evidenced in Fig. \ref{I-V2}, where
we report the current normalized to its saturation value as a function of
the applied voltage. For the sake of comparison, in this figure we also plot
the $I-V$ characteristics at $T=0$ which, with the used normalization,
becomes a universal function, independent of statistics, given by: 
\begin{equation}
\left. \frac{I}{I_{s}}\right| _{T=0}=\left\{ 
\begin{array}{c}
\frac{qV}{\mu }\text{ for }\frac{qV}{\mu }<1 \\ 
1\text{ for }\frac{qV}{\mu }>1
\end{array}
\right. \text{.}  \label{IVT0}
\end{equation}
As can be seen in the figure, the difference between the different
statistics amounts to a few percent of the value in the transition region.

\subsection{Noise properties}

As stated before, in the present structure current fluctuations originate
solely from the randomness of carrier injection from the contacts. The
carrier injection into the ballistic region must be, of course, consistent
with the statistics of the quasi-particles. For the case of contacts at
equilibrium, this consistency is satisfied by assuming that the low
frequency current spectral density in the energy range $(\varepsilon
,\varepsilon +d\varepsilon )$, is given by $s_{I}(\varepsilon -\mu
)d\varepsilon $, with\cite{Isakov99,Greiner99} 
\begin{equation}
s_{I}(\varepsilon -\mu )=2G_{0}k_{B}T\frac{\partial f_{g}(\varepsilon -\mu )%
}{\partial \mu }\text{.}  \label{sI}
\end{equation}
This expression is reminiscent of that for the variance of the mean
occupation number at equilibrium, $\left\langle \delta
f_{g}^{2}\right\rangle =k_{B}T\frac{\partial f_{g}(\varepsilon -\mu )}{%
\partial \mu }$, whose validity for FES was proved in Ref.[%
\onlinecite{Rajagopal95}]. Note that, for the case of Fermi-Dirac statistics
one has $s_{I}=2G_{0}f_{FD}(1-f_{FD})$, which shows the familiar term $%
f(1-f) $, while for FES in general it is $s_{I}=2G_{0}$ $f_{g}\left(
1-gf_{g}\right) \left( 1+(1-g)f_{g}\right) $. In all the cases with $g>0$,
in the zero temperature limit the low frequency noise vanishes since $%
f_{g}\rightarrow 1/g\theta (\mu -\varepsilon )$, and hence the transport
becomes fully coherent.

The current fluctuations can be computed as:\cite{Isakov99} 
\begin{equation}
S_{I}=\int_{0}^{+\infty }\left[ s_{I}(\varepsilon -\mu )+s_{I}(\varepsilon
+qV-\mu )\right] d\varepsilon \text{,}  \label{SI}
\end{equation}
where we have used that the channel is perfectly ballistic and transmitting
and that, because of strongly degenerate conditions, long range Coulomb
interaction can be neglected. After substituting Eq.(\ref{sI}) in Eq.(\ref
{SI}), one obtains 
\begin{eqnarray}
S_{I} &=&2G_{0}k_{B}T\left[ f_{g}(-\mu )+f_{g}(qV-\mu )\right]  \nonumber \\
&=&2k_{B}T(G^{eq}+G)\text{.}  \label{SI_G}
\end{eqnarray}
We note that, independently of carrier statistics, at thermal equilibrium
one always has: 
\begin{eqnarray}
S_{I}^{eq} &=&4k_{B}TG_{0}f_{g}(-\mu )  \nonumber \\
&=&4k_{B}TG^{eq}\text{,}  \label{SIeq}
\end{eqnarray}
thus recovering Nyquist theorem, as it should be. In addition, under current
saturation conditions one can neglect the contribution from one of the
contacts and obtain 
\begin{eqnarray}
S_{I}^{s} &=&2G_{0}k_{B}Tf(-\mu )  \label{SIs_f} \\
&=&2k_{B}TG^{eq}\text{,}  \label{SI_s}
\end{eqnarray}
which corresponds to half of Nyquist thermal noise. Similarly to the case of
the $I-V$ characteristics, the effects of the statistics are noticeable in
both the linear and saturation regimes through $G^{eq}$. In addition, the
transition region between these two regimes also shows a slight but
significant dependence on the statistics, as evidenced in Fig. \ref{SI-V}.
Here we plot the low frequency spectral density of current fluctuations as
given in Eq.(\ref{SI_G}) normalized to its saturation value, as a function
of the applied potential, for strongly degenerate conditions.

It is worth noting that, by using Eqs.(\ref{Geq_deg}), (\ref{Is_deg}) and (%
\ref{SI_s}), the saturation value of the current spectral density under
strongly degenerate conditions can be written as: 
\begin{equation}
\left( S_{I}^{s}\right) ^{\text{deg}}=\frac{2G_{0}k_{B}T}{g}=2q\left( \frac{%
k_{B}T}{\mu }\right) I_{s}\text{,}  \label{SIs_deg}
\end{equation}
which can be interpreted as shot noise suppressed by the degeneracy factor $%
k_{B}T/\mu $.

A convenient figure of merit of shot noise is the Fano factor, $\gamma $,
defined as 
\begin{equation}
\gamma =\frac{S_{I}}{2qI}\text{.}  \label{Fano}
\end{equation}
In Fig. \ref{Fano-V} we plot the Fano factor for different statistics as a
function of the applied potential under strongly degenerate conditions.
Here, at the lowest and highest potentials the Fano factor is independent of
statistics. More precisely, in the former limit it decreases inversely with
the applied voltage according to Nyquist relation as: 
\begin{equation}
\gamma =2\frac{k_{B}T}{qV}\text{ for }qV/\mu \ll 1\text{.}  \label{Fano_lowV}
\end{equation}
By contrast, in the latter limit the Fano factor is found to saturate taking
the value 
\begin{equation}
\gamma =\frac{k_{B}T}{\mu }\text{ for }qV/\mu \gg 1\text{.}
\label{Fano_highV}
\end{equation}
Between these limits, the Fano factor displays a transition region which
depends slightly on the carrier statistics. We note that the value reached
by $\gamma $ at the highest voltages is always less than one, thus
corresponding to suppressed shot noise, independently of carrier statistics.

In a certain sense, the picture emerging from the previous analysis is close
to that presented in Ref. [\onlinecite{Isakov99}]. Indeed, in both cases the
system displays thermal noise at low bias and suppressed shot noise at
higher bias, both limits being independent of carrier statistics, and with
the effect of carrier statistics becoming only noticeable in the transition
region between low and high bias voltages. The difference between the
present results and those of Ref.[\onlinecite{Isakov99}] is that in our
case, being the channel ballistic, the shot noise is reached under current
saturation conditions when $qV>\mu $ and displays a suppression factor equal
to $k_{B}T/\mu $, while in Ref.[\onlinecite{Isakov99}] being the channel
non-ballistic, the shot noise is reached for $qV<\mu $ and displays a
suppression factor $1-t$, where $t$ is the transmission of the channel.

At first sight the previous results are somewhat surprising. Because of the
sensitivity of shot noise to carrier correlations, one would have expected a
strong dependence of the Fano factor on carrier statistics. In particular,
one would have expected some evidence of the attractive or repulsive nature
of different statistics in the results obtained for strongly degenerate
conditions. However, the weak dependence of the noise properties on the
statistical parameter found above is a direct consequence of the strongly
degenerate conditions assumed. This conclusion will be better clarified in
the next section where the case of weakly degenerate conditions is
investigated.

\section{Weakly degenerate systems}

\label{Weakly} Under weakly degenerate conditions (i.e. $\mu <3k_{B}T$), to
evidence the effects of carrier statistics on the noise properties it
suffices to consider the current saturation regime. This limit offers the
advantage that the effects of long range Coulomb interaction on fluctuations
can be neglected, thus allowing us to resume results obtained in the
previous section.

The general expression for the saturation current is given in Eq.(\ref{Is}).
We note that under non-degenerate conditions (i.e. $\mu <-3k_{B}T$) one has $%
I_{s}^{n-\deg }=I_{0}f_{g}(-\mu )=I_{0}\exp (\mu /k_{B}T)$, which is
independent of $g$ as it should be. For intermediate values of the chemical
potential, the values of $I_{s}$ follow Eq.(\ref{Is}), thus interpolating
between the non-degenerate and the strongly degenerate values, given in Eq.(%
\ref{Is_deg}).

Furthermore, under saturation conditions the current spectral density is
given in general by Eq.(\ref{SIs_f}), and the Fano factor under saturation
by 
\begin{equation}
\gamma _{S}=\frac{f_{g}(-\mu )}{\ln \left( \frac{1+(1-g)f_{g}(-\mu )}{%
1-gf_{g}(-\mu )}\right) }\text{,}  \label{Fanos_f}
\end{equation}
where we have used Eqs.(\ref{Is}), (\ref{SIs_f}) and (\ref{Fano}). The
previous results can be written explicitly in terms of the saturation
current, by combining Eqs.(\ref{Is}), (\ref{SIs_f}) and \ref{Fanos_f}. One
then obtains the following expressions 
\begin{eqnarray}
S_{I}^{s} &=&\frac{2G_{0}k_{B}T}{g+\left[ exp(I_{s}/I_{0})-1\right] ^{-1}}%
\text{,}  \label{SIs_Is} \\
\gamma _{s} &=&\frac{I_{0}/I_{s}}{g+\left[ exp(I_{s}/I_{0})-1\right] ^{-1}}%
\text{.}  \label{Fanos_Is}
\end{eqnarray}
The above expressions give $S_{I}^{s}$ and $\gamma _{s}$ as a function of $%
I_{s}$, $q$ (through $G_{0}$ and $I_{0}$) and $g$, and can be used to obtain
direct information on particle statistics. When the injection of carriers
occurs under strongly degenerate conditions (i.e., $\mu /k_{B}T>3$, $%
I_{s}/I_{0}\gg 1$) we recover the results obtained in the previous section,
that is: 
\begin{eqnarray}
\left( S_{I}^{s}\right) ^{\text{deg}} &=&\frac{2G_{0}k_{B}T}{g}=2q\left( 
\frac{k_{B}T}{\mu }\right) I_{s}\text{,}  \label{SIs_deg2} \\
\left( \gamma _{s}\right) ^{\text{deg}} &=&\frac{k_{B}T}{\mu }\text{,}
\label{Fanos_deg}
\end{eqnarray}
thus corresponding to suppressed shot noise, with a Fano factor independent
of carrier statistics given by $k_{B}T/\mu $. Furthermore, under
non-degenerate conditions (i.e $\mu /k_{B}T<-3$, $I_{s}/I_{0}\ll 1$) we can
approximate Eqs. (\ref{SIs_Is}) and (\ref{Fanos_Is}) by 
\begin{eqnarray}
\left( S_{I}^{s}\right) ^{\text{non-deg}} &=&2qI_{s}\text{,}
\label{SIs-ndeg} \\
\left( \gamma _{s}\right) ^{\text{non-deg}} &=&1\text{,}  \label{Fanos_ndeg}
\end{eqnarray}
thus recovering full shot noise. This result is a direct consequence of the
fact that the distribution function for non-degenerate conditions is the
Maxwell-Boltzmann one. Obviously, in this limit the results are independent
of carrier statistics.

In the transition region between weakly and strongly degenerate conditions
(i.e., $|\mu |<3k_{B}T$) one must use the full expressions given in Eqs.(\ref
{SIs_f}) and (\ref{Fanos_f}), or alternatively in Eqs.(\ref{SIs_Is}) and (%
\ref{Fanos_Is}). Figure \ref{Fano_s} reports the Fano factor under current
saturation conditions versus the chemical potential, as obtained from Eq.(%
\ref{Fanos_f}), with the values of the distribution function being computed
from Eq.(\ref{f_alfa}). Here, the Fano factor is found to depend
significantly on particle statistics. In particular, in the transition
region $|\mu |<3k_{B}T$ the ''attractive'' or ''repulsive'' nature of the
exclusion statistics becomes manifest by the fact that the Fano factor takes
values larger or smaller than one, respectively. By analyzing Eq.(\ref
{Fanos_f}) it can be shown that for $0<g<1/2$ one has 
\begin{eqnarray}
\gamma _{s} &>&1\text{ \quad for }\mu <\mu _{c}\text{,}  \label{enhan1} \\
\gamma _{s} &<&1\text{ \quad for }\mu >\mu _{c}\text{,}  \label{enhan2}
\end{eqnarray}
where $\mu _{c}$ is the value of the chemical potential for which $\gamma
_{s}=1$, and that can be calculated for each statistics from Eqs.(\ref
{Fanos_f}). On the other hand, for $1/2\leq g\leq 1$ we obtain 
\begin{equation}
\gamma _{s}<1\text{ \quad for all }\mu \text{.}  \label{supp}
\end{equation}

Therefore, for carriers following FES with $0<g<1/2$ we have proved their
possibility to display enhanced shot noise ($\gamma _{s}>1$), thus
evidencing the positive correlation (bunching) induced by the exclusion
statistics. Remarkably, these positive correlations are only manifested
under weakly degenerate conditions and when $\mu <\mu _{c}$. Under strongly
degeneracy conditions, the shot noise is always suppressed since the system
tends to a coherent transport, which implies noiseless conditions.

By contrast, for carriers following FES with $1/2\leq g\leq 1$ we have
obtained always suppressed shot noise ($\gamma _{s} < 1$), thus evidencing
the negative correlation (anti-bunching) induced by the exclusion statistics.

\section{Conclusions}

\label{Conclusions} We have studied the noise properties of perfect
ballistic, one-dimensional channels with carriers satisfying FES. We have
found that the attractive or repulsive nature of the carrier statistics can
be really appreciated only in the case of systems which are under weakly
degenerate conditions, that is with a chemical potential $\mu $ in the
contacts satisfying the condition $|\mu |<3k_{B}T$. Then, under current
saturation regime the Fano factor is found to display enhanced shot noise ($%
\gamma _{s}>1$), or suppressed shot noise ($\gamma _{s}<1$) depending on
whether the statistical parameter is in the range $0<g<1/2$ , or in the
range, $1/2\leq g\leq 1$, respectively. These results show that for
particles with $0<g<1/2$ the statistics tends to bunch the carriers, while
for particles with $1/2\leq g\leq 1$ it tends to antibunch them.

In the remaining range of values of the chemical potential ($|\mu |>3k_{B}T$%
) we have found that the effects of the statistics on the noise properties
are less pronounced. For the case of strongly degenerate systems ($\mu
>3k_{B}T$) the information on the particle statistics is mostly contained in
the linear region of the $I-V$ characteristics, and to a less extent in the
transition region between linear and current saturation conditions. In
particular, independently of carrier statistics the system always displays
suppressed shot noise under current saturation conditions, with a Fano
factor $\gamma _{s}=k_{B}T/\mu <1$. On the other hand, for the case of
non-degenerate systems, ($\mu <-3k_{B}T$), the distribution function is well
approximated by the Maxwell-Boltzman distribution, and hence the results are
independent of the carrier statistics. In this case, under current
saturation conditions the system displays full shot noise, $\gamma _{s}=1$.

Present results suggest that to evidence the nature of the FES one should
consider situations in which the system is under weakly degenerate
conditions. This conclusion is expected to remain basically valid also for
one dimensional channels with transmission less than one.

\section{Acknowledgments}

This work has been performed within the framework of the EC Improving Human
Research Potential program through contract No. HPMF-CT-1999-00140.

\begin{figure}[tbp]
\caption{Renormalized $I-V$ characteristics of the ballistic conductor for
different statistics under strongly degenerate conditions. $I_s$ is the
saturation current. For comparison the results for T=0 are also reported. }
\label{I-V2}
\end{figure}

\begin{figure}[tbp]
\caption{Current spectral density at low frequency of the ballistic
conductor as function of the applied bias for different statistics under
strongly degenerate conditions. The current spectral density is normalized
to its value under current saturation conditions.}
\label{SI-V}
\end{figure}

\begin{figure}[tbp]
\caption{Fano factor of the ballistic conductor as a function of the applied
bias under strongly degenerate conditions for different statistics}
\label{Fano-V}
\end{figure}

\begin{figure}[tbp]
\caption{Fano factor of the ballistic conductor under current saturation
conditions as a function of the chemical potential for different statistics. 
$k_BT$ is the thermal energy}
\label{Fano_s}
\end{figure}



\begin{references}
\bibitem{Haldane91}  F.D.M. Haldane, Phys. Rev. Lett. {\bf 67}, 937 (1991).

\bibitem{Isakov94a}  S. B. Isakov, Int. J. Mod. Phys. A {\bf 9}, 2563 (1994).

\bibitem{Ha94}  Z.N.C. Ha, Phys. Rev. Lett. {\bf 73}, 1574 (1994); {\bf 74},
620 (1995).

\bibitem{Murthy94}  M.V.N. Murthy and S. Shankar, Phys. Rev. Lett. {\bf 73},
3331 (1994).

\bibitem{Wu95}  Y.S. Wu and Y. Tu, Phys. Rev. Lett. {\bf 75}, 890 (1995).

\bibitem{Hatsugai96}  Y. Hatsugai, M. Kohmoto, T. Koma, and Y. S. Wu, Phys.
Rev. B {\bf 54}, 5358 (1996).

\bibitem{Pepin99}  C. P\'{e}pin and A. M. Tsvelik, Phys. Rev. Lett. {\bf 82}%
, 3859 (1999).

\bibitem{Bhaduri96}  R.K. Bhaduri, M.V.N. Murthy, and M.K. Srivastava, Phys.
Rev. Lett. {\bf 76}, 165 (1996).

\bibitem{Srivastava99}  M.K. Srivastava, R.K. Bhaduri, J. Law, and M.V.N.
Murthy, cond-mat/9902158.

\bibitem{Dasnieres94}  A. Dasni\`{e}res de Veigy, S. Ouvry, Phys. Rev. Lett. 
{\bf 72}, 600 (1994).

\bibitem{Chen95}  W. Chen and Y.J. Ng, Phys. Rev. B, {\bf 51}, 14479 (1995).

\bibitem{Su96}  W.P. Su, Y.S. Wu and J. Yang, Phys. Rev. Lett. {\bf 77},
3423 (1996).

\bibitem{Yu97}  Y. Yu, W. Zheng and Z. Zou, Phys. Rev. B {\bf 56}, 13279
(1997).

\bibitem{Wu98}  Y.S. Wu, Y. Yu, Y. Hatsugai and M. Kohmoto, Phys. Rev. B 
{\bf 57}, 9907 (1998).

\bibitem{Elburg98}  R.A.J. van Elburg and K. Schoutens, Phys. Rev. B {\bf 58}%
, 15704 (1998). See also, W. Zheng and Y. Yu, Phys. Rev. Lett. {\bf 79},
3242 (1997); Y. Yu, Phys. Rev. B {\bf 61}, 4465 (2000).

\bibitem{Isakov94b}  S. Isakov, Mod. Phys. Lett. B {\bf 8}, 319 (1994).

\bibitem{Wu94}  Y.S. Wu, Phys. Rev. Lett. {\bf 73}, 922 (1994).

\bibitem{Rajagopal95}  A.K. Rajagopal, Phys. Rev. Lett. {\bf 74}, 1048
(1995).

\bibitem{Isakov96}  S.B. Isakov, D.P. Arovas, J. Myrheim, and A.P.
Polychronakos, Phys. Lett. A {\bf 212}, 299 (1996).

\bibitem{Isakov97}  S.B. Isakov, S. Mashkevich, Nucl. Phys. B {\bf 504}, 701
(1997).

\bibitem{Iguchi97}  K. Iguchi, Phys. Rev. Lett. {\bf 78}, 3233 (1997).

\bibitem{Su98}  G. Su and M. Suzuki, E. Phys. J. B, {\bf 5}, 577 (1998).

\bibitem{Rego99}  L.G.C. Rego and G. Kirczenow, Phys. Rev. B {\bf 59}, 13080
(1999).

\bibitem{Krive99}  I.V. Krive and E.R. Mucciolo, Phys. Rev. B {\bf 60}, 1429
(1999).

\bibitem{Isakov99}  S.B. Isakov, T. Martin and S. Ouvry, Phys. Rev. Lett. 
{\bf 83}, 580 (1999).

\bibitem{Greiner99}  A. Greiner, L. Reggiani and T. Kuhn, Physica B {\bf 272}%
, 75 (1999).

\bibitem{Gomila00}  G. Gomila and L. Reggiani, Semicond. Sci. Technol. {\bf %
15}, 829 (2000).

\begin{figure}[tbp]
\caption{$I-V$ characteristics of the ballistic conductor for different
statistics under strongly degenerate conditions. Here, $%
I_{0}=(2s+1)qk_{B}T/h $ and $\mu $ is the chemical potential}
\label{I-V}
\end{figure}
\end{references}
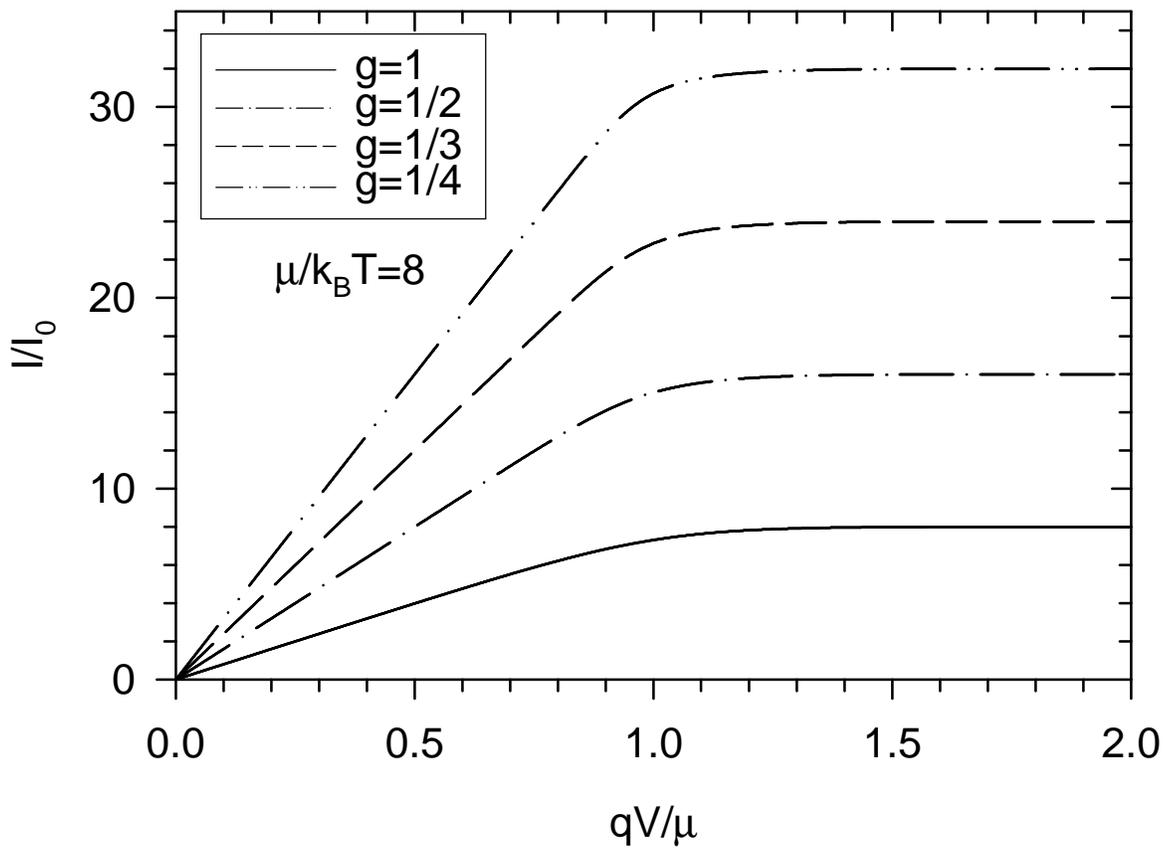
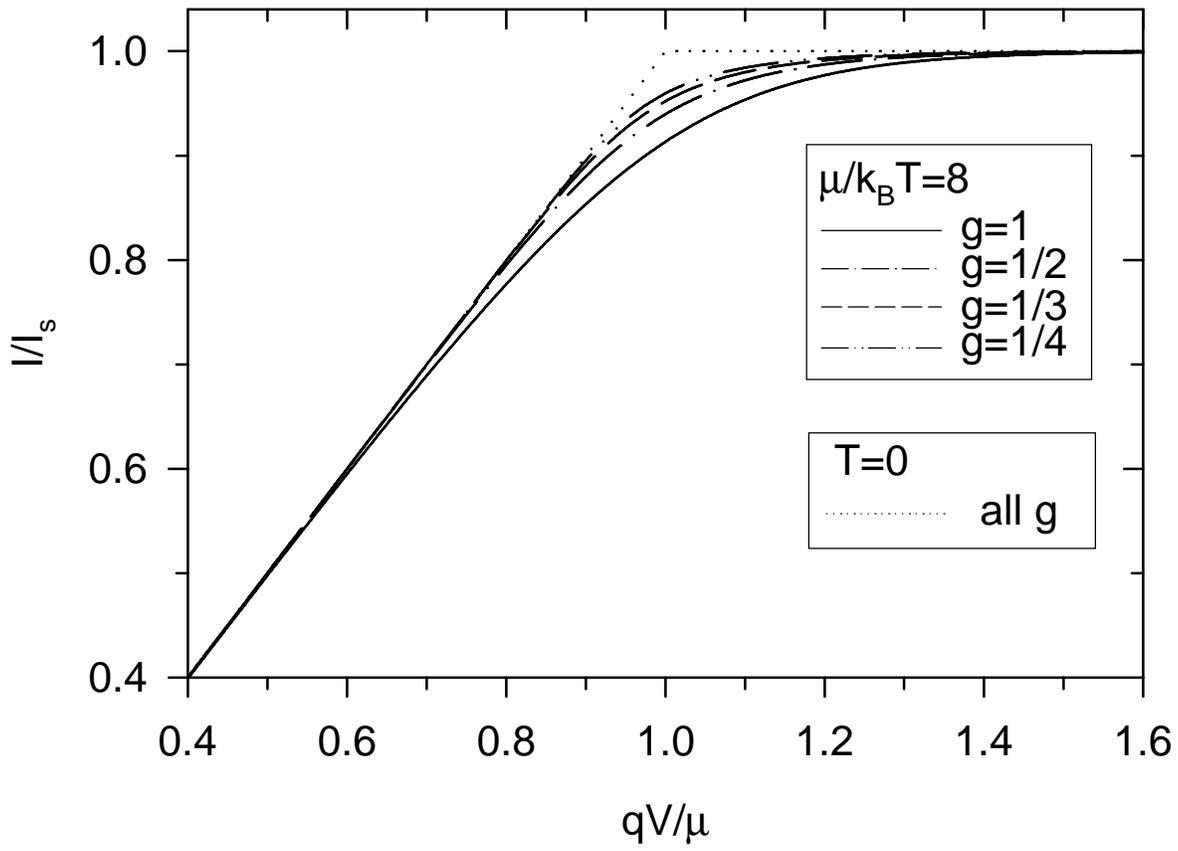
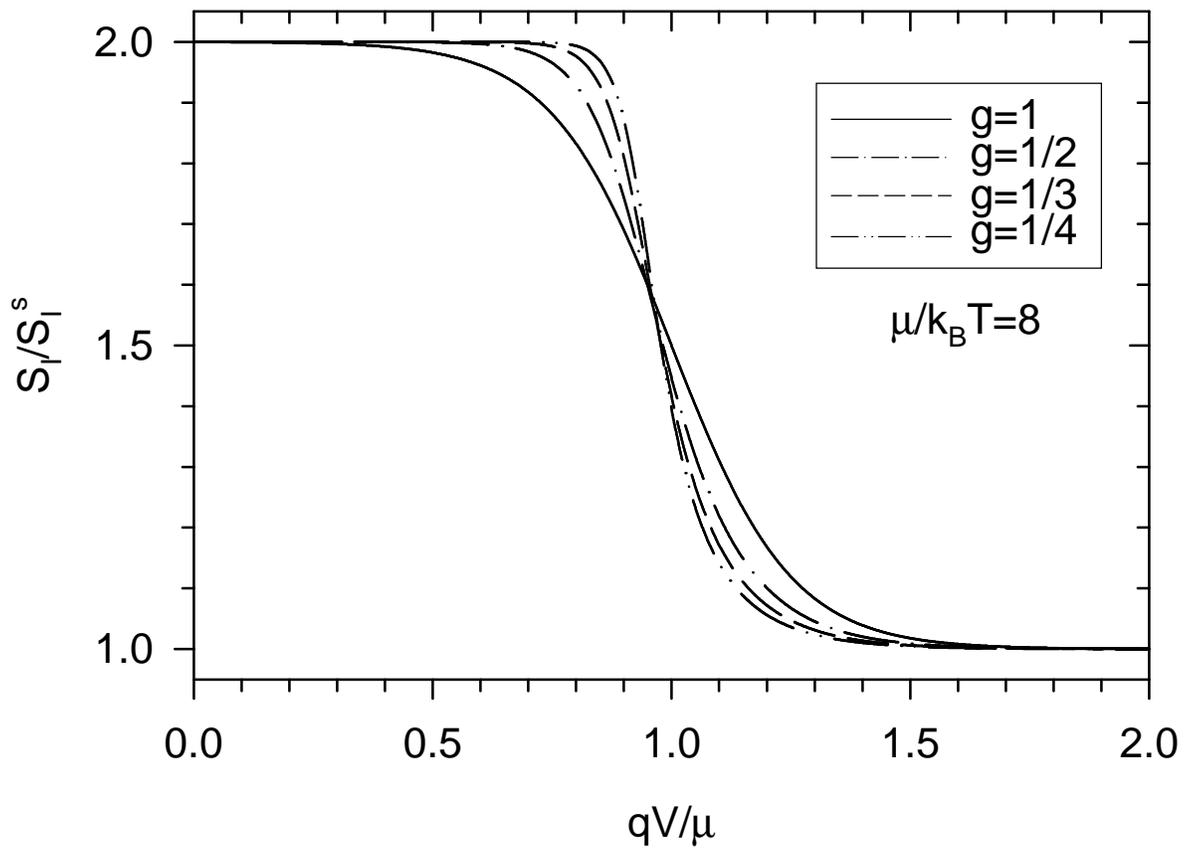
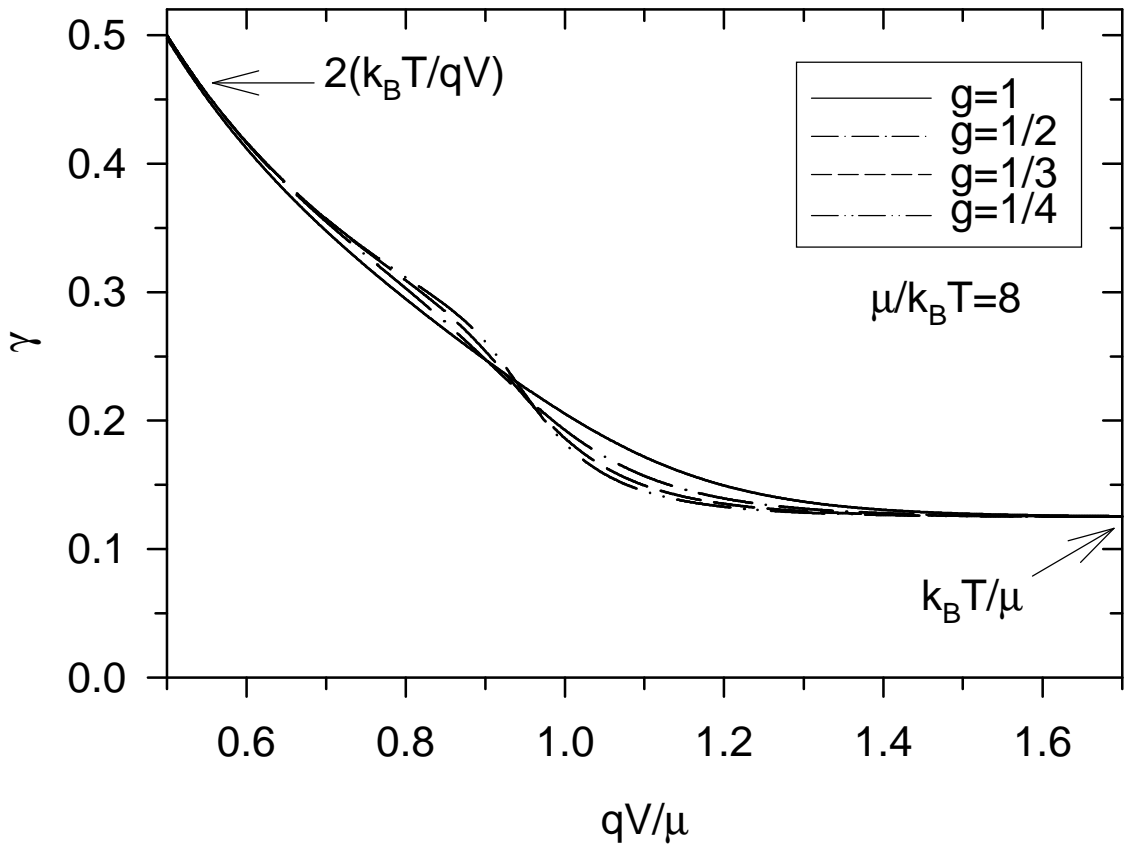
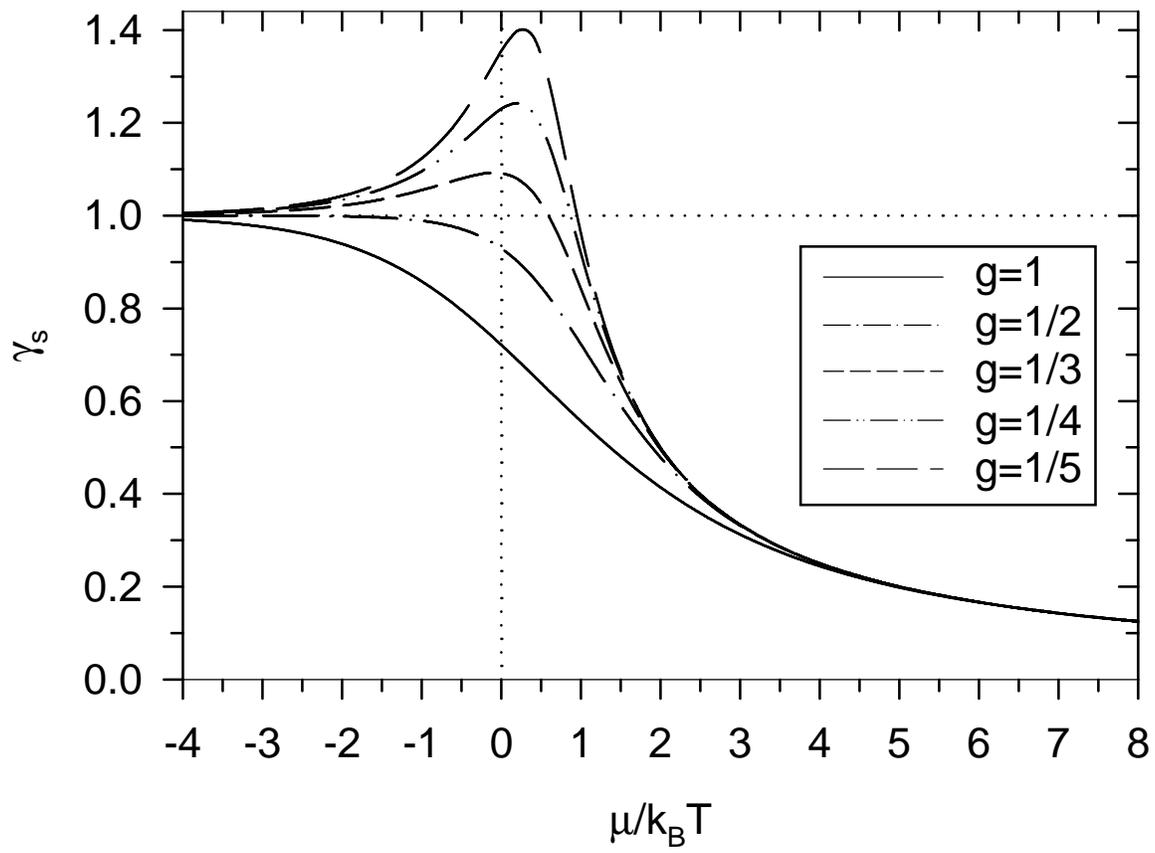
\end{document}